\documentstyle[11pt,Moriond,epsfig]{article}
\begin{document}
\vspace*{4,cm}
\date{}
\title{\uppercase{Strangeness enhancement in heavy ion collisions}}
\author{\uppercase {Krzysztof Redlich$^{1,2}$ and Ahmed Tounsi$^3$}}
\address{
$^1$Gesellschaft f\"ur Schwerionenforschung, D-64291 Darmstadt,
Germany\\
$^2$Institute of Theoretical Physics, University of
Wroc\l aw, PL-50204 Wroc\l aw, Poland\\
$^3$ LPTHE, Universit\'e Paris 7, 2 pl Jussieu, F--75251 Cedex 05}
\maketitle\abstracts{
We argue  that the main features of baryon and anti-baryon
enhancement observed by  the WA97 collaboration can be described
using canonical formulation of strangeness conservation. Within
this formulation strangeness enhancement could be larger at lower
collision energies. }
\section{Introduction}
Ultrarelativistic heavy ion collisions provide a unique
opportunity to study the properties of  highly excited hadronic
matter under extreme conditions of high density and temperature
\cite{qm,satz,stoc,stachel,heinz1}.  From the analysis of rapidity
distribution of protons and of their transverse energy measured in
158 A GeV/c Pb+Pb collisions an estimate of the initial condition
 leads to energy density of 2-3
GeV/fm$^3$ and a baryon density  of the order of 0.7/fm$^3$.
Lattice QCD at vanishing baryon density suggest that the phase
transition from confined to the quark-gluon plasma (QGP) phase
appears at the temperature $T_c=173\pm 8$ MeV which corresponds to
the critical energy density \cite{karsch} $\epsilon_c\sim 0.6\pm
0.3$ GeV/fm$^3$. One could thus conclude that the required initial
conditions for quark deconfinement are already reached in heavy
ion collisions at SPS energy. It will be discussed to what extend
the composition of the final-state hadrons, in particular their
strangeness content can be considered as a probe of quark
deconfinement in the initial state.
\section{Strangeness in quark-gluon plasma}

It was  argued that enhanced production of strange particles could
be a signal of QGP formation in heavy ion collisions
\cite{rafelski1}. In the QGP the production and equilibration of
strangeness is very efficient due to the a large gluon density and
a low energy threshold for dominant QCD processes of  $s\bar s$
production. In the hadronic system the higher threshold for
strangeness production  was expected
 to make the strangeness yield considerably
smaller and the equilibration time much longer. (Recently,  it was
argued  \cite{rapp,greiner} and shown in terms of a microscopic
transport approach \cite{cassing}  that multi-mesonic reactions
could, however,  accelerate the equilibration time of antibaryons
in high density hadronic medium.)

From these strangeness QGP  characteristics one expects: a {\it
global strangeness enhancement\/}
 measured  by the total number of produced $<s\bar s>$ quarks per
 participant $A_{part}$ or per produced light quarks
$<u\bar u +d\bar d>$ which should increase from pp, pA to AA collisions
and {\it enhancement of multistrange (anti-)baryons}
in central AA collisions, with respect to proton induced
reactions.

Large strangeness content of the QGP plasma should be reflected in
a very specific hierarchy of multistrange baryon enhancement:
$E_{\Lambda}  < E_{\Sigma} < E_{\Omega}$. This hierarchy is
observed by the WA97 and NA57  collaboration
\cite{andersen,carrer} which measure
 the yield per participant in Pb+Pb relative to p+Pb and p+Be collisions.
Indeed the enhancement pattern of the (anti)hyperon yields is seen
to increase with strangeness content of the particle and there is
a saturation of this enhancement for $A_{part} > 100$. Recent
results  of the NA57 collaboration are showing in addition an
abrupt  change of anti-cascade enhancement for lower centrality.
Similar behavior  was previously seen on the
 $K^+$ yield measured by the NA52 experiment in Pb-Pb
collisions \cite{kabana}. These results are very interesting as
they might be interpreted as an indication of the onset of new
dynamics. However, a more detailed experimental study and
theoretical understanding are still required here. It is, e.g, not
clear  what is the origin of different centrality dependence of
$\Xi$ and $\bar\Xi$ as well as inconsistences   between the NA52
and the WA97 data.

A number of different mechanisms
\cite{soff,bravina,vance,bleicher,capella,lin} was considered   to
describe the magnitude of the enhancement and centrality
dependence of (multi)strange baryons measured by the WA97.
Microscopic transport models make it clear that the WA97 data  can
not be explained by pure final state  hadronic interactions. Only
the combination of the formers with an additional pre-hadronic
mechanisms like baryon junction processes, color ropes or color
flux tubes overlap can partly explain the enhancement pattern and
the magnitude for the most central collisions. However, the
detailed centrality dependence is still not well reproduced within
the microscopic models. An alternative description of multistrange
particle production was formulated in terms of macroscopic thermal
models with \cite{braun,hamieh} and without \cite{letessier}
chemical equilibrium. In equilibrium the parameters are
temperature $T$, baryon  and strangeness chemical potentials. In
non-equilibrium one also uses two other parameters $\gamma_q$ for
$u$ and $d$ quarks and $\gamma_s$ for $s$ quark. These parameters
measure deviations from full equilibrium. The fits of reference
\cite{letessier} with more parameters are more precise, but fail
to reproduce the $\Omega$ and $\bar \Omega$ yields. Besides, the
hadronization temperature  is found lower ($T_f\sim 145$ MeV) than
the one found in equilibrium ($T_f\simeq 170$ MeV). This led to
the development of  sudden hadronization model with super-cooling
\cite{letessier}.

However the equilibrium thermal model \cite{braun,becattini2}
describes all existing WA97 and NA49 data. The model was
formulated  in the grand canonical (GC) ensemble with a partition
function  which contains  the contributions of most hadrons and
resonances and preserves  the baryon number, strangeness and
charge conservation. The particle ratios depends only on two
parameters;  temperature $T$ and baryon chemical potential
$\mu_B$. With $T_f\sim 170$ MeV, corresponding to the  energy
density $\epsilon_f\sim 0.6$GeV/$fm^3$, and with  the baryon
chemical potential $\mu_B\sim 270$MeV, corresponding to baryon
density of 0.16GeV/$\rm fm^3$, the statistical model reproduces
the experimental data. Furthermore this model  is also
 consistent \cite{xu,redlich,new} with the recent data of the STAR
collaboration in Au+Au collisions at RHIC \cite{harris}. The main
difference with SPS, as expected because of higher collision
energy with less stopping, is a smaller value ($\mu_B\sim 50$ MeV)
of baryon chemical potential. The freezeout temperatures in
central A+A collisions at the SPS and RHIC coincide within errors
\footnote{ At RHIC a precise determination of $T_f$ is not yet
possible as fits with $160<T_f<200$   give similar value of
$\chi^2$. Data like, e.g, $\Omega /\pi$ or $\Xi /\pi$ ratios are
necessary to a better determination of $T_f$. \cite{new}} with the
critical temperature from lattice QCD.  This could  indicate that
all particles are originating from deconfined medium  and that the
chemical composition of the system is established during
hadronization \cite{stoc,stachel,heinz1}.

\begin{figure}[htb]
\vskip -1.cm
\begin{minipage}[t]{80mm}
\includegraphics[width=17.5pc, height=16.5pc]{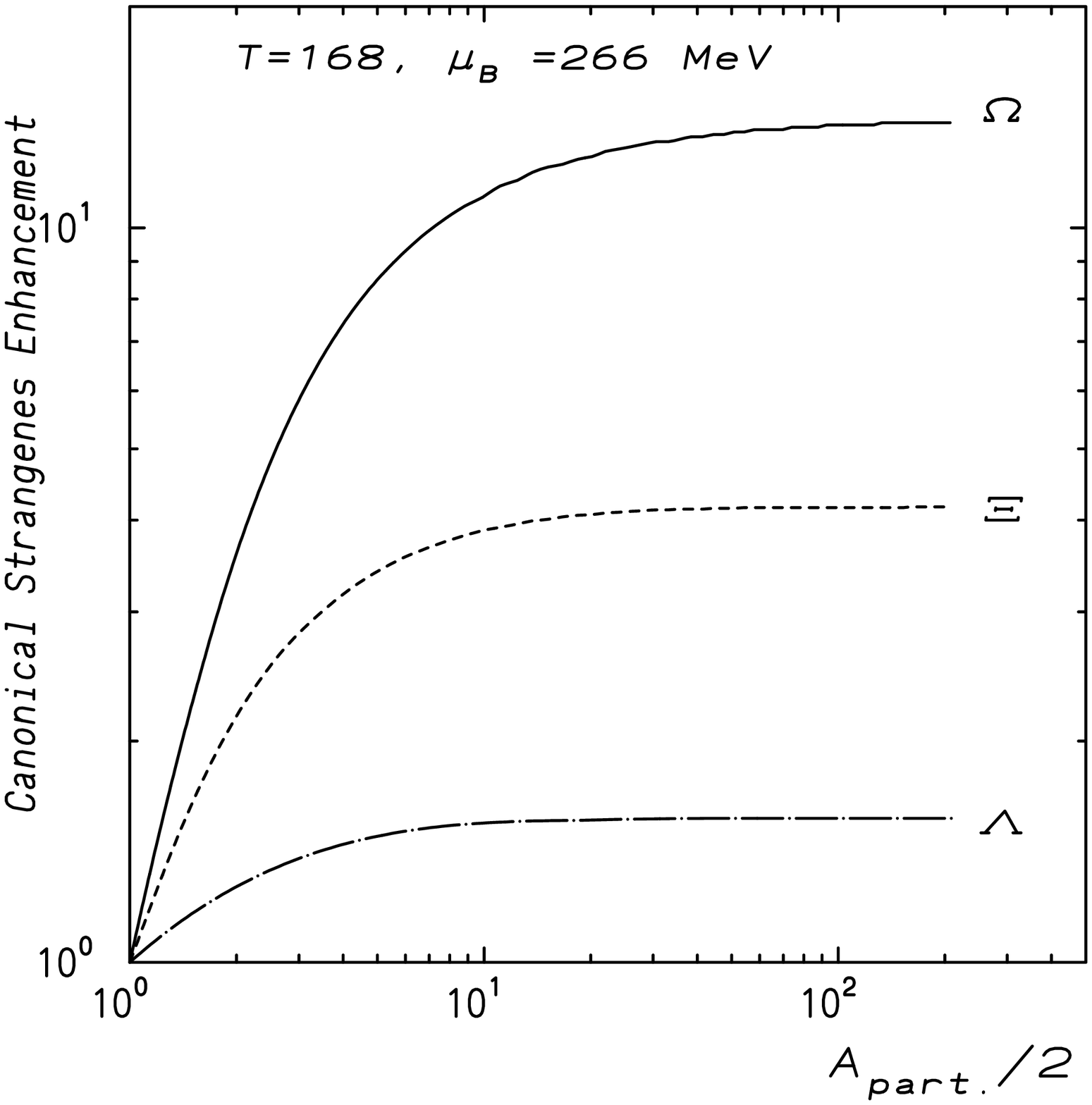}\\
\vskip -1. true cm \caption{Particle multiplicities per
participant normalized to its value in p+A system as a function of
$A_{part.}$ calculated  in statistical model in (C)
ensemble$^{18}$.\hfill}
\end{minipage}
\hspace{\fill}
\begin{minipage}[t]{75mm}
\includegraphics[width=17.5pc, height=16.5pc]{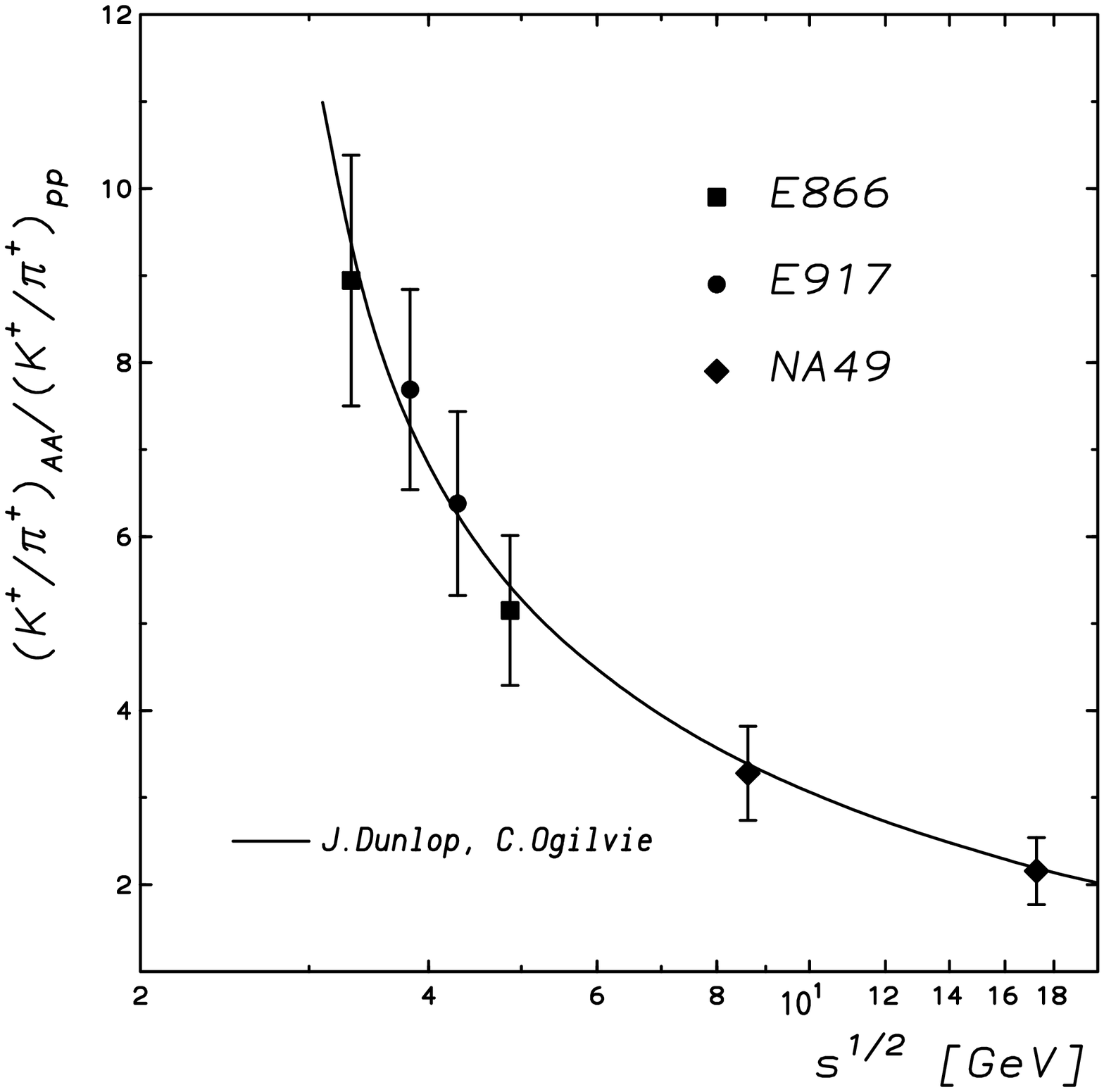}\\
\vskip -1. true cm \caption{ $K^+/\pi^+$ ratio at midrapidity in
A+A relative to p+p collisions. For the compilation of data see
refs. [33,34]. \hfill}
\end{minipage}
\end{figure}

\section{Canonical description of stangeness enhancement}
The enhancement of strange(anti)-baryons reported by the WA97 and
its variation with centrality may be explained in  canonical
formulation of statistical thermal model. What the WA97 measures
is the ratio of strange (anti-)baryon yields in the large Pb-Pb
system and in the small p-Be or p-Pb system. For small systems
with rare  production of strange particles strangeness
conservation must be implemented locally on an event-by-event
basis \cite{ko}, i.e, canonical (C) ensemble for strangeness
conservation must be used. The (C) ensemble is relevant in the
statistical description of particle production in low energy
 heavy ion \cite{cleymans1}, or high energy hadron-hadron or
$e^+e^-$ reactions \cite{b1} as well as in peripheral heavy ion
collisions at SPS \cite{hamieh}. The exact conservation of quantum
numbers, that is the canonical approach, is known to severely
reduce the phase-space available for particle productions
\cite{hagedorn}.

Fig.1 shows the multiplicity per participant of $\Omega , \Xi ,$
and $\Lambda$   relative to its value in a small system with only
two participants. One sees that the statistical model in (C)
ensemble reproduce the basic features of WA97 data: the
enhancement pattern and  enhancement saturation for large
$A_{part}$ indicating here  that  (GC) limit is reached. The
quantitative comparison \cite{hamieh} of the model with
experimental  data requires an additional assumption on the
variation of $\mu_B$ with centrality to account for  larger value
of $\bar B/B$ ratios in p+A than in Pb+Pb collisions
\cite{andersen}. However, an abrupt change of enhancement seen in
the NA57 data for $\bar\Xi$ is  very unlikely to be reproduced in
terms of this approach.

One of the consequences  of the model is that the enhancement
pattern seen in Fig. 1 should not be only a feature of the SPS
data. In terms of statistical  model strangeness enhancement and
enhancement pattern should be also present in heavy ion collisions
at lower energies and should be even more pronounced. This is in
contrast e.g. to UrQMD predictions which are showing increasing
enhancement with beam energy \cite{bleicher}. In Fig.2 we show a
compilation of the data on $K^+/\pi^+$ ratio in A+A relative to
p+p collisions. This double ratio could be referred to as a
strangeness enhancement factor. The enhancement is indeed seen to
be larger  at the smaller beam energy and is smoothly decreasing
towards higher energy. If strangeness enhancement is indeed of
thermal origin then  similar behavior is also expected for
multistrange baryons. This could put in question \cite{last} the
observed strangeness enhancement  from p+p to A+A as an {\it
appropriate characteristics} to signal deconfinement. To possibly
observe QGP from the strangeness composition of the fireball one
should rather search for non-monotonic behavior of strange
particle species versus centrality or collision energy.
\section*{Acknowledgments}
\hspace*{\parindent} We  acknowledge  stimulating discussions with
R. Baier,   P. Braun-Munzinger, B. Friman,  M. Gazdzicki, W.
N\"orenberg, H. Oeschler, H. Satz, and R. Stock.

\end{document}